\documentclass[10pt]{article}

\usepackage{graphicx}
\usepackage{amsmath}
\usepackage{amssymb}

\numberwithin{equation}{section}
\newtheorem{thm}{Theorem}[section]
\newenvironment{prf}{\noindent {\it Proof} \ }{\hfill $\Box$}
\newenvironment{rmk}{\noindent {\it Remark} \ }{}

\newtheorem{defn}[thm]{Definition}
\newtheorem{cor}[thm]{Corollary}
\newcommand{\eqa}{\begin{eqnarray}}
\newcommand{\eeqa}{\end{eqnarray}}
\newcommand{\beq}{\begin{equation}}
\newcommand{\eeq}{\end{equation}}
\newcommand{\nn}{\nonumber}
\newcommand{\pp}{\partial}
\newcommand{\VE}{\varepsilon}

\newcommand{\lm}{\lambda}

\newcommand\pal{\partial}
\begin{document}
\parskip 6pt
\hoffset -1.8cm

\title{A 2-Component Generalization of the Camassa-Holm Equation and Its Solutions }
\author{Ming Chen, Si-Qi Liu, Youjin Zhang\\
{\small Department of Mathematical Sciences, Tsinghua
University}\\
{\small Beijing 100084, P.R. China}}
\date{}
\maketitle

\begin{abstract}
An explicit reciprocal transformation between a 2-component
generalization of the Camassa-Holm equation, called the 2-CH
system, and the first negative flow of the AKNS hierarchy is
established, this transformation enables one to obtain solutions
of the 2-CH system from those of the first negative flow of the
AKNS hierarchy. Interesting examples of peakon and multi-kink
solutions of the 2-CH system are presented.
\vskip 0.5 cm
\noindent {\bf Mathematics Subject Classifications(2000)}. 35Q53, 37K35\newline
{\bf Key words}: Camassa-Holm equation, AKNS hierarchy, reciprocal transformation
\end{abstract}

\section{Introduction}

The Camassa-Holm equation, which was derived physically as a shallow water wave equation
by Camassa and Holm in \cite{CH, CHH}, takes the form
\beq\label{CH}
u_t+\kappa\, u_x-u_{xxt}+3 u u_x=2 u_x u_{xx}+u u_{xxx}
\eeq
where $u=u(x,t)$ is the fluid velocity in the $x$ direction and the constant $\kappa$ is related to the critical
shallow water wave speed. The subscripts $x, t$ of $u$ denote the partial derivatives of the function $u$ w.r.t. $x, t$,
for example, $u_t=\pal_t u,\ u_{xxt}=\pal_t\pal_x\pal_x u$, similar notations will be
used frequently later in this paper.
This equation
first appeared in the work of Fuchssteiner and Fokas \cite{FF} on their theory of hereditary symmetries
of soliton equations.
As it was shown by Camassa and Holm, equation (\ref{CH}) shares most of the important properties of an
integrable system of KdV type, for example, the existence of Lax pair formalism, the
bihamiltonian structure, the multi-soliton solutions and the applicability of
the inverse scattering method to its initial value problem. When $\kappa=0$, the Camassa-Holm equation
(\ref{CH}) has a peculiar property that its soliton
solutions become piecewise smooth and have corners at their crests, such solutions are
weak solutions of (\ref{CH}) and are called ``peakons''. Since the works of Camassa and Holm,
this equation has become a well known example of integrable systems and has been studied
from various point views in, for example, \cite{AG, Alb, Cons-1, hone-1, Fo, Fu, hone,johnson,mckean-1, mckean-2, schiff}
and references therein.

We consider in this paper the following 2-component generalization of the Camassa-Holm equation
\begin{eqnarray}
&&m_t+u\,m_x+2\,m\,u_x-\rho\,\rho_x=0,\label{ch-2}\\
&&\rho_t+(\rho\,u)_x=0.\label{ch-3}
\end{eqnarray}
Here $m=u-u_{xx}+\frac12\kappa$. Under the constraint $\rho=0$ this system is reduced to the Camassa-Holm
equation (\ref{CH}). Such a generalization is based on the following consideration. We note the fact
that both bihamiltonian structures of the Camassa-Holm hierarchy and that of the KdV hierarchy
are deformations of the following bihamiltonian structure
of hydrodynamic type
\begin{eqnarray}
&&\{u(x),u(y)\}_1=\delta'(x-y),\nn\\
&&\{u(x),u(y)\}_2=u(x)\delta'(x-y)+\frac12\,u(x)'\delta(x-y).\label{z8}
\end{eqnarray}
This fact also implies that the dispersionless limit of the Camassa-Holm hierarchy coincides with that of the
KdV hierarchy. It was shown in \cite{DLZ, LZ1} that deformations of the bihamiltonian structure (\ref{z8})
that depend polynomially on the variables $u_x, u_{xx},\dots$ are uniquely characterized,
up to Miura type transformations,
by a function
$c(u)$, this function is called the central invariant of the deformed bihamiltonian structure. For
the KdV hierarchy the central invariant is a nonzero constant while for the
Camassa-Holm hierarchy the central invariant is given by a nonzero constant multiplied by $u$, so their
bihamiltonian structures
are representatives of two different classes of deformations of (\ref{z8}). One of the main features
of the integrable hierarchies that correspond to bihamiltonian structures with constant central invariants
is the existence of tau functions \cite{DZ}, this property no longer holds true for integrable hierarchies
that correspond to bihamiltonian structures with nonconstant central invariants.
In fact, many of the well known integrable hierarchies of evolutionary PDEs
with one spatial variable possess bihamiltonian structures that are deformations of
bihamiltonian structure of hydrodynamic type with constant central invariants, and the existence
of tau functions plays an important role in the study of these integrable systems. The Camassa-Holm
hierarchy is an exceptional example of integrable systems which does not possess tau functions.
Now let us consider the following bihamiltonian structure of hydrodynamic type
\begin{eqnarray}
&&\{w_1(x), w_1(y)\}_1=\{w_2(x), w_2(y)\}_1=0,\nn\\
&&\{w_1(x), w_2(y)\}_1=\delta'(x-y).\label{z10-a}\\
&&\{w_1(x), w_1(y)\}_2=2 \delta'(x-y),\nn\\
&&\{w_1(x), w_2(y))\}_2=w_1(x) \delta'(x-y)+w_1'(x) \delta(x-y),\nn\\
&&\{w_2(x), w_2(y)\}_2=\left[w_2(x)\pal_x+\pal_x w_2(x))\right] \delta(x-y).\label{z10}
\end{eqnarray}
It was shown in \cite{LZ1} that the deformation with constant central invariants
$c_1=c_2=\frac1{24}$ leads to the nonlinear Schr\"odinger hierarchy which can be converted to the AKNS
hierarchy \cite{akns} after an appropriate transformation, and the one with central invariants
$c_1=\frac1{24}\,{(w_1+2\sqrt{w_2})^2}, c_2=\frac1{24}\,{(w_1-2\sqrt{w_2})^2}$ leads to a
bihamiltonian integrable hierarchy with (\ref{ch-2}),
(\ref{ch-3}) as the first nontrivial flow under the change of coordinates
\beq
w_1=2(u-u_x),\quad w_2=-\rho^2+(u-u_x)^2,
\eeq
the rescaling $t\mapsto -t$ and the Galilean transformation $x\mapsto x-\frac12\kappa\,t,\ t\mapsto t,\
 u\mapsto u-\frac12 \kappa,\
\rho\mapsto \rho$.
So from the point of view of deformations of bihamiltonian structures of hydrodynamic type the systems (\ref{CH})
and
(\ref{ch-2}), (\ref{ch-3}) have the same property, i.e., both of their bihamiltonian structures have nonconstant central invariants.
Note that the bihamiltonian structure of the system (\ref{ch-2}), (\ref{ch-3}), as it was shown in \cite{LZ1},
is obtained from (\ref{z10-a}), (\ref{z10})
by the addition of the deformation
term $-\delta''(x-y)$ to the bracket $\{w_1(x), w_2(y)\}_1$. We will call the system (\ref{ch-2}), (\ref{ch-3}) the
2-component Camassa-Holm (2-CH) system henceforth. This system was also derived independently by Falqui \cite{falqui}
by using the bihamiltonian approach.

The main result of the present paper is the establishment of a reciprocal transformation between the 2-CH system and
the first negative flow of the AKNS hierarchy.  Recall that
the Camassa-Holm equation (\ref{CH}) has a similar relation with the first negative flow of the KdV hierarchy,
the corresponding reciprocal transformation (also called a hodograph transformation) was found by
Fuchssteiner in \cite{Fu}.  Several attempts have been made to obtain solutions
of the Camassa-Holm equation (\ref{CH}) from that of the first negative flow of the KdV hierarchy by using
this reciprocal transformation in, for example, \cite{Cons-1, hone, johnson,
LZJ1, LZJ2, schiff}. However, since the
inverse of this reciprocal transformation involves the solving of a nonlinear ODE of second order,
only particular solutions like the multi-soliton
solutions  were obtained in explicit forms by using this approach.
The advantage of the reciprocal transformation between the 2-CH system and
the first negative flow of the AKNS hierarchy is that it gives an explicit correspondence between solutions of
these two systems, this correspondence is presented in Theorems \ref{fuv}, \ref{fx} and Theorems \ref{qrf},
\ref{fqr}. In Sec.\,\ref{sec-2} and Sec.\,\ref{sec-3}
we first give the construction of the
reciprocal transformation, then in Sec.\,\ref{sec-4}
we show some interesting examples of solutions of the 2-CH system that are obtained from solutions of the
first negative flow of the AKNS hierarchy by using the reciprocal transformation, they
include peakon and  multi-kink solutions.

\section{A reciprocal transformation for the 2-CH system }\label{sec-2}

The 2-CH system is equivalent to the compatibility conditions of the following
linear systems:
\eqa
&&\phi_{xx}+\left(-\frac14+m\,\lambda-\rho^2\,\lambda^2\right)\phi=0,\label{lax-a}\\
&&\phi_t=-(\frac{1}{2\lm}+u)\,\phi_x+\frac{u_x}2\phi.\label{lax-b}
\eeqa Here $m=u-u_{xx}+\frac12\kappa$. Since the term
$\frac12\kappa$ can be canceled by a Galileo transformation, we
assume $\kappa=0$ in this and the next section. The linear
equation (\ref{lax-a}) is known as the Schr\"odinger spectral
problem with energy dependent potential. Antonowicz and Fordy
considered the more general spectral problem \beq\label{fordy}
\phi_{xx}+(c+u_1\,\lm+\dots+u_n\,\lm^n)\phi=0 \eeq and associated
to it $n+1$ compatible local Hamiltonian structures in \cite{fordy1}.
Here $c$ is a given constant. In \cite{fordy1} it was also shown
that (\ref{fordy}) can be transformed to the spectral problem \beq
\psi_{xx}+(v_0+v_1\,\lm+\dots+v_{n-1}\,\lm^{n-1})\psi=\lm^n\psi.\label{alonso}
\eeq We will use similar transformations below in order to relate
the 2-CH system with the first negative flow of the AKNS
hierarchy. The relations of the spectral problem (\ref{alonso})
and its generalizations to multi-Hamiltonian structures and
integrable systems were studied in \cite{alonso, fordy2, fordy3},
and in the particular case of $n=2$ in \cite{jaulent}. The
bihamiltonian structure of the 2-CH system (\ref{ch-2}),
(\ref{ch-3}) was also given
 in \cite{fordy1} from the spectral problem (\ref{fordy}) with $n=2$. In \cite{DLZ} the bihamiltonian structures
related to the generalized spectral problems of \cite{fordy3} were considered from the point of view
of deformations of bihamiltonian structures of  hydrodynamic type, their central invariants are in general non-constants.

Since when $\rho$ vanishes, the 2-CH system (\ref{ch-2}), (\ref{ch-3}) degenerates to the
Camassa-Holm equation (\ref{CH}), we assume hereafter $\rho \ne 0$.
Equation (\ref{ch-3}) shows that the $1$-form
\beq
\omega=\rho\,dx-\rho\,u\,dt
\eeq
is closed, so it define a reciprocal transformation $(x,t)\mapsto (y,s)$ by the relation
\begin{equation}\label{rt-1}
dy=\rho\,dx-\rho\,u\,dt,\quad ds=dt,
\end{equation}
and we have
\begin{equation}\label{zh9}
\frac{\pp}{\pp x}=\rho\,\frac{\pp}{\pp y},\quad \frac{\pp}{\pp t}=\frac{\pp}{\pp s}-\rho\,u\,\frac{\pp}{\pp y}.
\end{equation}
Denote $\varphi=\sqrt{\rho}\,\phi$, then the spectral problem (\ref{lax-a}), (\ref{lax-b}) is
converted to the following one
\eqa
&&\varphi_{yy}+\left(-\lm^2+P\,\lm+Q\right)\varphi=0,\label{nlax-a}\\
&&\varphi_s+\frac{\rho}{2\lm}\varphi_y-\frac{\rho_y}{4\lm}\varphi=0,\label{nlax-b}
\eeqa
where
\beq
P=\frac{m}{\rho^2},\quad
Q=-\frac1{4\rho^2}-\frac{\rho_{yy}}{2\rho}+\frac{\rho_y^2}{4\rho^2}.\label{tr-2}
\eeq

Now let's consider the isospectral problem (\ref{nlax-a}) (\ref{nlax-b}). The compatibility conditions read
\begin{eqnarray}
&&P_s=\rho_y, \label{ed-1} \\
&&Q_s+\frac12\rho\,P_y+P\,\rho_y=0, \label{ed-2} \\
&&\frac12\rho\,Q_y+Q \rho_y+\frac14\rho_{yyy}=0. \label{ed-3}
\end{eqnarray}
By integrating the third equation (\ref{ed-3}) and comparing the resulting equation with (\ref{tr-2}) we obtain
\begin{equation}\label{ed-4}
\rho^2\,Q+\frac12\rho\,\rho_{yy}-\frac14\rho_y^2=C=-\frac14.
\end{equation}
From the equation (\ref{ed-1}) we know that
there exists a function $f(y,s)$ such that
\begin{equation} \label{tr-3}
P=\frac{\pp f(y,s)}{\pp y},\ \rho=\frac{\pp f(y,s)}{\pp s}.
\end{equation}
Substituting the expressions of $P,Q,\rho$ that are given by (\ref{tr-3}) and the second formula of
(\ref{tr-2}) into the equation (\ref{ed-2}) we arrive at the following equation for $f$:
\begin{equation}\label{ef}
\frac{f_{ss}}{2f_s^3}+f_yf_{ys}-\frac{f_{ss}f_{ys}^2}{2f_s^3}+\frac{f_{ys}f_{yss}}{2f_s^2}
+\frac12f_sf_{yy}+\frac{f_{ss}f_{yys}}{2f_s^2}-\frac{f_{yyss}}{2f_s}=0
\end{equation}

\begin{thm}\label{fuv}
Let $f$ be a solution of the equation (\ref{ef}), and
\begin{equation}
u=f_yf_s^2+\frac{f_{ss}f_{ys}}{f_s}-f_{yss},\quad \rho=f_s. \label{zh6}
\end{equation}
If $x(y,s)$ is a solution of the following system of ODEs:
\begin{equation}
\frac{\pp x}{\pp y}=\frac1{\rho},\quad  \frac{\pp x}{\pp s}=u, \label{zh7}
\end{equation}
then $(u(y,t),\rho(y,t),x(y,t))$ is a parametric solution of the 2-CH system (\ref{ch-2}), (\ref{ch-3}).
\end{thm}
\begin{rmk}
We say that the triple $(u(y,t),\rho(y,t),x(y,t))$ is a parametric solution of the 2-CH system if
the functions ${\bar{u}}(x,t)=u(y(x,t),t),\,  {\bar{\rho}}(x,t)=\rho(y(x,t),t)$ satisfy the system (\ref{ch-2}),
(\ref{ch-3}), here $y=y(x,t)$ is the inverse function of $x=x(y,t)$. For simplicity, we will use the
same symbol $u, \rho$ to denote the functions $u(y(x,t),t),\,  \rho(y(x,t),t)$ as functions of $x$ and  $t$.
\end{rmk}

\begin{prf}
Due to the definition of the reciprocal transformation, we only need to verify the validity of the equation
\beq
u-u_{xx}=m=\rho^2 P=f_s^2 f_y.\label{zh5}
\eeq
Denote by $E$ the l.h.s of equation (\ref{ef}). By using the definition (\ref{zh6}) of the function $u$
we obtain through a straightforward computation
$$u-\rho\,(\rho\,u_y)_y-f_s^2 f_y+2f_s^3 E_y+4 f_s^2 f_{ys}E=0$$
which yields (\ref{zh5}).
The theorem is proved.
\end{prf}

\begin{defn} A function $f=f(y,s)$ is called a primary solution of
the 2-CH system (\ref{ch-2}), (\ref{ch-3}) if it satisfies the equation (\ref{ef}).
\end{defn}

Given a solution $(u(x,t), \rho(x,t))$ of the 2-CH system (\ref{ch-2}), (\ref{ch-3}), the formulae
(\ref{tr-2}) and (\ref{tr-3}) determines a primary solution $f(y,s)$, we call it the primary solution
that is associated to $(u(x,t), \rho(x,t))$. On the other hand, any primary solution $f(y,s)$
yields a solution of the 2-CH system in a parametric form through the formulae (\ref{zh6}), (\ref{zh7}).
In the next theorem it will be shown that from a primary solution $f(y,s)$ one can construct another
solution of the 2-CH system. This solution is still in parametric form, however,
in this case the function $x=x(y,s)$ is given explicitly in terms of $f(y,s)$ without the need of integration.

Due to (\ref{zh6}), we can rewrite the equations of (\ref{zh7}) in the form
\beq
\frac{\pp x}{\pp y}=\frac1{f_s}, \quad
\frac{\pp x}{\pp s}=f_yf_s^2+\frac{f_{ss}f_{ys}}{f_s}-f_{yss}.\label{xsfy}
\eeq
By using the equations (\ref{tr-3}), (\ref{zh7}) and the first equation of (\ref{tr-2}) we also have
\beq
\frac{\pp f}{\pp s}=\frac1{x_y}, \quad
\frac{\pp f}{\pp y}=\frac{m}{\rho^2}=\frac{u-u_{xx}}{\rho^2}=x_sx_y^2+\frac{x_{yy}x_{sy}}{x_y}-x_{syy}.\label{fyxs}
\eeq
The similarity of these equations suggests the existence of
a duality between solutions of the equation (\ref{ef}). Indeed,
we have the following theorem.

\begin{thm}\label{fx}
Let $f(y,s)$ be a solution of the equation (\ref{ef}). Define the functions $x=x(y,s), u=u(y,s), \rho=\rho(y,s)$ by
\begin{equation}\label{zh8}
x=f(s,y),\quad u=\frac{\pp x}{\pp s},\quad \frac1{\rho}=\frac{\pp x}{\pp y}.
\end{equation}
Then $(u(y,t),\rho(y,t),x(y,t))$ is a parametric solution of the 2-CH system (\ref{ch-2}), (\ref{ch-3}).
\end{thm}
\begin{prf}
Substiting
$
u=\frac{\pp x}{\pp s},\ \frac1{\rho}=\frac{\pp x}{\pp y}
$
into the equations (\ref{ch-2}), (\ref{ch-3}) we know, by using the relation (\ref{zh9}), that
$(u(y,t),\rho(y,t),x(y,t))$ gives a solution to the 2-CH system if and only if the function $x(y,s)$ satisfies
\begin{equation}\label{exx}
x_{ss}+\frac{2x_sx_{ys}}{x_y}+\frac{x_{yy}}{x_y^4}-\frac{x_{ys}^2x_{yy}}{x_y^4}+
\frac{x_{yss}x_{yy}}{x_y^3}+\frac{x_{ys}x_{yys}}{x_y^3}-\frac{x_{yyss}}{x_y^2}=0.
\end{equation}
This equation follows immediately from the fact that the function $f(y,s)$ satisfies (\ref{ef}).
The theorem is proved.
\end{prf}

Let us note that for the parametric solution (\ref{zh8}), the associated primary solution ${\bar f}(y,s)$ that is
determined by the formulae (\ref{tr-2}), (\ref{tr-3}) is in general different from the original primary
solution $f(y,s)$. This procedure yields a B\"acklund transformation $f(y,s)\mapsto {\bar f}(y,s)$ for the
equation (\ref{ef}). We will consider in detail such a class of B\"acklund transformation for the equation
(\ref{ef}) and the 2-CH system in another publication. In the next section, we will show how to construct
primary solutions of the 2-CH system from solutions of the first negative flow of the AKNS hierarchy.

\section{Relations to the first negative flow of the AKNS hierarchy}\label{sec-3}

The AKNS spectral problem is given by
\begin{equation}\label{lax-3}
\left(\begin{array}{c} \phi_1 \\ \phi_2 \end{array}\right)_y=
\left(\begin{array}{rr} \lm & -q \\ r & -\lm \end{array}\right)
\left(\begin{array}{c} \phi_1 \\ \phi_2 \end{array}\right).
\eeq
The first negative flow of the ANKS hierarchy is equivalent to the compatibilty conditions of
(\ref{lax-3}) with the linear system
\beq\label{lax-4}
\left(\begin{array}{c} \phi_1 \\ \phi_2 \end{array}\right)_s=
\frac1{4\lm}\left(\begin{array}{rr} a & b \\ c & -a \end{array}\right)
\left(\begin{array}{c} \phi_1 \\ \phi_2 \end{array}\right).
\end{equation}
This flow can be represented in the form
\begin{eqnarray}
&&q_s=\frac12\, b,\quad r_s=\frac12\, c, \label{ak-1}\\
&&b_y=2\,a\,q,\quad c_y=2\,a\,r, \label{ak-2}\\
&&a_y+b\,r+c\,q=0.\label{ak-3}
\end{eqnarray}
Substitute the equation (\ref{ak-2}) into (\ref{ak-3}), we obtain
\begin{equation}\label{kk}
a^2+b\,c=\VE^2,
\end{equation}
where $\VE$ is a constant. We assume that $\VE \ne 0$.

\begin{thm}\label{qrf}
Let $(a,b,c,q,r)$ be a solution of the equations (\ref{ak-1})--(\ref{ak-3}) with $\VE^2=1$,
then any function $f(y,s)$ satisfying
\begin{equation}\label{ff-1}
2a=b\,e^{-f}-c\,e^f
\end{equation}
gives a primary solution of the 2-CH system.
\end{thm}
\begin{prf}
Assume that the function $f=f(y,s)$ satisfies (\ref{ff-1}). Let us first prove the following formula:
\beq
f_y=q\,e^{-f}+r\,e^f \label{ff-3}
\eeq
Due to (\ref{ff-1}) and (\ref{ak-2}), (\ref{ak-3}) we have
\begin{eqnarray}
0&=&(2a-b\,e^{-f}+c\,e^f)_y=2\,a_y-b_y\,e^{-f}+c_y\,e^f+(b\,e^{-f}+c\,e^f)f_y \nn \\
&=&-2(b\,r+c\,q)-2\,a\,q\,e^{-f}+2\,a\,r\,e^f+(b\,e^{-f}+c\,e^f)f_y \nn \\
&=&-2(b\,r+c\,q)-(b\,e^{-f}-c\,e^f)(q\,e^{-f}-r\,e^f)+(b\,e^{-f}+c\,e^f)f_y \nn \\
&=&(b\,e^{-f}+c\,e^f)(f_y-q\,e^{-f}-r\,e^f)\label{zh1}
\end{eqnarray}
The formula (\ref{ff-3}) then follows from (\ref{zh1}) and the fact that
\begin{equation}\label{ff-2}
b\,e^{-f}+c\,e^f=\sqrt{(b\,e^{-f}-c\,e^f)^2+4\,b\,c}=\sqrt{4(a^2+b\,c)}=\pm 2\VE \ne 0
\end{equation}
Differentiating both sides of (\ref{ff-3}) w.r.t. $y$ and $s$ we obtain respectively
\begin{eqnarray}
f_{yy}&=&(q\,e^{-f}+r\,e^f)_y=q_y\,e^{-f}+r_y\,e^f-(q\,e^{-f}-r\,e^f)f_y \nn \\
&=&q_y\,e^{-f}+r_y\,e^f-(q\,e^{-f}-r\,e^f)(q\,e^{-f}+r\,e^f) \nn \\
&=&q_y\,e^{-f}+r_y\,e^f-q\,e^{-2f}+r\,e^{2f} \label{ff-4}
\end{eqnarray}
\begin{eqnarray}
f_{ys}&=&(q\,e^{-f}+r\,e^f)_s=q_s\,e^{-f}+r_s\,e^f-(q\,e^{-f}-r\,e^f)f_s \nn \\
&=&\frac12b\,e^{-f}+\frac12c\,e^f-(q\,e^{-f}-r\,e^f)f_s \label{ff-5}
\end{eqnarray}
For any solution $(\phi_1=\phi_1(y,s;\lm), \phi_2=\phi_2(y,s;\lm))$ of the systems (\ref{lax-3}), (\ref{lax-4}) define
\begin{eqnarray}
&&\varphi=e^{-\frac{f}2}\phi_1+e^{\frac{f}2}\phi_2, \label{zh10} \\
&& P=f_y,\ \rho=f_s,\label{zh2}\\
&&Q=-\frac34q^2\,e^{-2f}-\frac12q\,r-\frac34r^2\,e^{2f}+\frac12q_y\,e^{-f}-\frac12r_y\,e^f.
\end{eqnarray}
By using equations (\ref{lax-3})-(\ref{ff-5}) and the fact $\varepsilon^2=1$, we can show through a straightforward and lengthy computation
that the functions $\varphi, P, Q, \rho$ satisfy the equations (\ref{nlax-a}), (\ref{nlax-b}) and (\ref{ed-4}).
To illustrate our computation, we give here the proof of the validity of equation (\ref{nlax-b}).
From (\ref{lax-3}), (\ref{lax-4}) and (\ref{ff-3}) it follows that
\begin{eqnarray}
\varphi_s&=&(e^{-\frac{f}2}\phi_1+e^{\frac{f}2}\phi_2)_s
=e^{-\frac{f}2}(\phi_1)_s+e^{\frac{f}2}(\phi_2)_s-(e^{-\frac{f}2}\phi_1-e^{\frac{f}2}\phi_2)\frac{f_s}2 \nn \\
&=&e^{-\frac{f}2}\frac{a\,\phi_1+b\,\phi_2}{4\lm}+e^{\frac{f}2}\frac{c\,\phi_1-a\,\phi_2}{4\lm}
-(e^{-\frac{f}2}\phi_1-e^{\frac{f}2}\phi_2)\frac{f_s}2 \nn \\
&=&\left(\frac{a+c\,e^f}{4\lm}-\frac{f_s}2\right)e^{-\frac{f}2}\phi_1+
\left(\frac{b\,e^{-f}-a}{4\lm}+\frac{f_s}2\right)e^{\frac{f}2}\phi_2
\end{eqnarray}
and
\begin{eqnarray}
\varphi_y&=&(e^{-\frac{f}2}\phi_1+e^{\frac{f}2}\phi_2)_y
=e^{-\frac{f}2}(\phi_1)_y+e^{\frac{f}2}(\phi_2)_y-(e^{-\frac{f}2}\phi_1-e^{\frac{f}2}\phi_2)\frac{f_y}2 \nn \\
&=&e^{-\frac{f}2}(\lm\,\phi_1-q\,\phi_2)+e^{\frac{f}2}(r\,\phi_1-\lm\,\phi_2)
-(e^{-\frac{f}2}\phi_1-e^{\frac{f}2}\phi_2)\frac{q\,e^{-f}+r\,e^f}2 \nn \\
&=&\left(\lm-\frac{q\,e^{-f}-r\,e^f}2\right)e^{-\frac{f}2}\phi_1+
\left(-\lm-\frac{q\,e^{-f}-r\,e^f}2\right)e^{\frac{f}2}\phi_2
\end{eqnarray}
By using these two formulae and (\ref{ff-5}), (\ref{zh2}) we obtain
\begin{eqnarray}
&&\varphi_s+\frac{\rho}{2\lm}\varphi_y-\frac{\rho_y}{4\lm}\varphi
=\varphi_s+\frac{f_s}{2\lm}\varphi_y-\frac{f_{ys}}{4\lm}\varphi \nn \\
&&=\frac{e^{-\frac{f}2}\phi_1-e^{\frac{f}2}\phi_2}{4\lm}\left(a-\frac{b}2\,e^{-f}+\frac{c}2\,e^f\right)=0.
\end{eqnarray}
The theorem then follows from the derivation of equation (\ref{ef}) that is given in the last section.
\end{prf}

\begin{rmk}
The main point in the proof of the above theorem is the establishment of the gauge transformation
(\ref{zh10}) between the isospectral problem (\ref{nlax-a}), (\ref{nlax-b}) and (\ref{lax-3}), (\ref{lax-4}).
For the positive flows of AKNS hierarchy, there also exist gauge transformations of form
$\varphi=e^{-\frac{f}2}\phi_1+e^{\frac{f}2}\phi_2$, where $f$ satisfies $f_y=q\,e^{-f}+r\,e^f$,
they establish correspondences between the positive flows of the AKNS hierarchy and the ones that are related to the
isospectral problems (\ref{nlax-a}).
\end{rmk}

The following theorem gives an explicit way of constructing a solution of the first negative flow of the AKNS hierarchy
from a primary solution of the 2-CH system.
\begin{thm}\label{fqr}
If $f$ is a primary solution of the 2-CH system (\ref{ch-2}), (\ref{ch-3}),
then we can construct a solution of the first negative flow of the AKNS hierarchy by the following formulae
\begin{equation}
q=\frac{e^f}2\left(f_y+\frac{\VE-f_{ys}}{f_s}\right),\ r=\frac{e^{-f}}2\left(f_y-\frac{\VE-f_{ys}}{f_s}\right),\
b=2\,q_s,\ c=2\,r_s,\ a=\frac{b\,e^{-f}-c\,e^f}2.
\end{equation}
where $\VE=1$ or $\VE=-1$.
\end{thm}
\begin{prf}
Since $f$ is a primary solution of the 2-CH system, we have $E=0$ where $E$ is defined as in the proof of
Theorem \ref{fuv}. Then by a straightforward computation, we obatin
\begin{equation}
b_y-2\,a\,q=2e^f\,E=0,\ c_y-2\,a\,r=2e^{-f}\,E=0,\ a^2+b\,c=\VE^2=1.
\end{equation}
The theorem is proved.
\end{prf}

By using the freedom in the choice of signs of the parameter $\VE$ we obtain the following corollary:
\begin{cor}\label{bkl}
We have the following two B\"acklund transformations for the equation (\ref{ef})
\begin{equation}
f \mapsto B_\VE f=f+\log\left(
\frac{f_{ss}f_{ys}+f_s^3f_y-f_sf_{yss}+\VE(f_{ss}-f_s^2)}{f_{ss}f_{ys}+f_s^3f_y-f_sf_{yss}+\VE(f_{ss}+f_s^2)}\right),\
\VE=1 \ {\rm or}\ -1.
\end{equation}
\end{cor}
\begin{prf}
Since $f$ is a primary solution of the 2-CH system, by using Theorem \ref{fqr} we obtain two solutions of the first
negative flow of the AKNS hierarchy, we denote them by
$(a_\VE,b_\VE,c_\VE,q_\VE,r_\VE), \VE=1, -1$. Due to Theorem \ref{qrf}, each of these two
solutions yields two primary solutions of the 2-CH system
\begin{equation}
f_{\VE,\gamma}=\log\frac{\gamma-a_\VE}{c_\VE}=\log\frac{b_\VE}{\gamma+a_\VE},\quad \VE, \gamma=\pm1.
\end{equation}
It is easy to see that $f_{\VE,\gamma}=f$ when $\gamma=\VE$. The corollary is then proved if we identify
$B_\VE f$ with $f_{\VE,-\VE}$.
\end{prf}

The B\"acklund transformations given by the above corollary can also be represented in terms of the
dependent variables $q, r$ of the first negative flow of the AKNS hierarchy, due to their long expressions
we do not present the formulae here. Instead, let us illustrate the procedure of obtaining solutions of the
system (\ref{ak-1})--(\ref{ak-3}) starting from the following trivial solution:
\begin{equation}\label{tsol1}
q=\sum_{i=1}^n e^{\zeta_i y+\frac{\gamma}{\zeta_i} s+\xi_i},\ r=0,\ b=2q_s,\ c=0,\ a=\gamma,
\end{equation}
where $n\in\mathbb{N}$,\ $\zeta_i, \xi_i, \gamma$ are arbitrary
constants with $\zeta_i$ non-vanishing and pairwise distinct. According
to Theorem \ref{qrf} we know that
$$f_0=\log\frac{b}{\gamma+a}=\log\frac{q_s}\gamma$$
is a primary solution of the 2-CH system. By using the above corollary, we obtain two solutions of the equation
(\ref{ef}), however, only one of them makes sense. By repeating this procedure, we obtain a sequence of primary
solutions $f_0,f_1,f_2,\dots$ of the 2-CH system. Then by using Theorem \ref{fuv} or Theorem \ref{fx} and Theorem
\ref{fqr}, we obtain a sequence of solutions of the 2-CH system and the first negative flow of the AKNS hierarchy.
Such class of solutions of the 2-CH system have very nice properties, we will study them in more detail
in the next section.

\section{Particular solutions of the 2-CH system}\label{sec-4}
We are now ready to present in this section some examples of solutions of the 2-CH system, they include the peakon and
multi-kink solutions.

{\bf Example 1.} Let us first try, without referring to the reciprocal transformation
constructed above, to find travelling wave solutions of the 2-CH system.
Assume  $u=h(x+v\,t),\rho=g(x+v\,t)$, where $v$ is a constant. Then the equation (\ref{ch-2}) and (\ref{ch-3}) become
\eqa\label{fff}
&&v(h'-h''')+3h\,h''-2h'\,h''-h\,h'''+\kappa\,h'-g\,g'=0,\\
&&v\,g'+g\,h'+h\,g'=0.
\eeqa
We can solve the second equation directly
\begin{equation}\label{gf} g=\frac{A}{v+h}\end{equation}
where A is a constant. Then the equation (\ref{fff}) becomes an ODE for $h$. This ODE can be solved by
a standard method. By carefully choosing the integration constant, we obtain the following solution
\begin{equation}\label{pk-1}
u=\chi-\sqrt{\chi^2-v^2},\ \rho=\sqrt{v\,K}\left(1+\sqrt{\frac{\chi-v}{\chi+v}}\right),\ {\rm where}
\ \chi=(v-K)\cosh(x+v\,t)+K,
\end{equation}
where $K=-4\kappa$. If $v>0,K>0$, this is a travelling peakon solution, see Figure 1.

{\bf Example 2.} The first negative flow of the AKNS hierarchy has an important reduction.
Under the assumption
\begin{equation}
q=-\frac{w_y}2,\ r=\frac{w_y}2,\ b=-\sinh w,\ c=\sinh w,\ a=\cosh w,
\end{equation}
it is reduced to the sinh-Gordon equation
\begin{equation}
w_{ys}=\sinh w.
\end{equation}

Now let us employ the results of the previous sections to obtain a stationary peakon solution of the 2-CH
system (\ref{ch-2}) and (\ref{ch-3})
with $\kappa=0$. Due to Theorem \ref{qrf}, a solution of the sinh-Gordon equation leads to a primary solution
of the 2-CH system
\begin{equation}
f(y,s)=\varepsilon \log\left(-\tanh\frac{w}2\right), {\rm where}\ \varepsilon=\pm1.
\end{equation}
Then the equations in (\ref{tr-3}) become
\begin{equation}
P=\varepsilon \frac{w_y}{w_{ys}},\ \rho=\varepsilon \frac{w_s}{w_{ys}}.
\end{equation}
By using Theorem \ref{fuv}, we find a parametric solution of the 2-CH system with $\kappa=0$
\begin{equation}\label{sg}
x(y,s)=\varepsilon \log w_s+x_0,\quad u(y,s)=\varepsilon \frac{w_{ss}}{w_s},\quad \rho=\varepsilon \frac{w_s}{w_{ys}}.
\end{equation}
where $x_0$ is an arbitrary constant.

Now let us choose a kink solution of the sinh-Gordon equation
\begin{equation}
w(y,s)=4 \tanh^{-1}\left(e^{p_1\, y+\frac{s}{p_1}+q_1}\right),
\end{equation}
with some constants $p_1\ne 0, q_1$ and substitute it into (\ref{sg}), we obtain a stationary solution of the 2-CH system
\begin{equation}
u=\frac{\sqrt{1+e^{2\varepsilon (x-x_0)}}}{p_1},\ \rho=\frac1{p_1\,\sqrt{1+e^{2\varepsilon (x-x_0)}}}
\end{equation}
Then it is easy to see that the following $u,\rho$
\begin{equation}\label{pk-2}
u=\frac{\sqrt{1+e^{-2|x-x_0|}}}{p_1},\ \rho=\frac1{p_1\sqrt{1+e^{-2|x-x_0|}}}
\end{equation}
is a stationary peakon solution of the 2-CH system, see Figure 1. Note that a peakon solution with constant
speed $-\frac12\kappa$ for the 2-CH system without the assumption of $\kappa=0$ can be obtained by the
Galilean transformation $x\mapsto\tilde x= x-\frac12\,\kappa\,t,\  t\mapsto \tilde t=t, \ u\mapsto \tilde u=u-\frac12\,\kappa,
\ \rho\mapsto \tilde \rho=\rho$.
If we choose $w$ as a multi-kink solution of the
sinh-Gordon equation, we will arrive at some very strange
solutions of the 2-CH system whose peaked points are also
stationary.

\begin{figure}
\begin{center} \small
$$
\begin{array}{cc}
\includegraphics[width=0.45\textwidth]{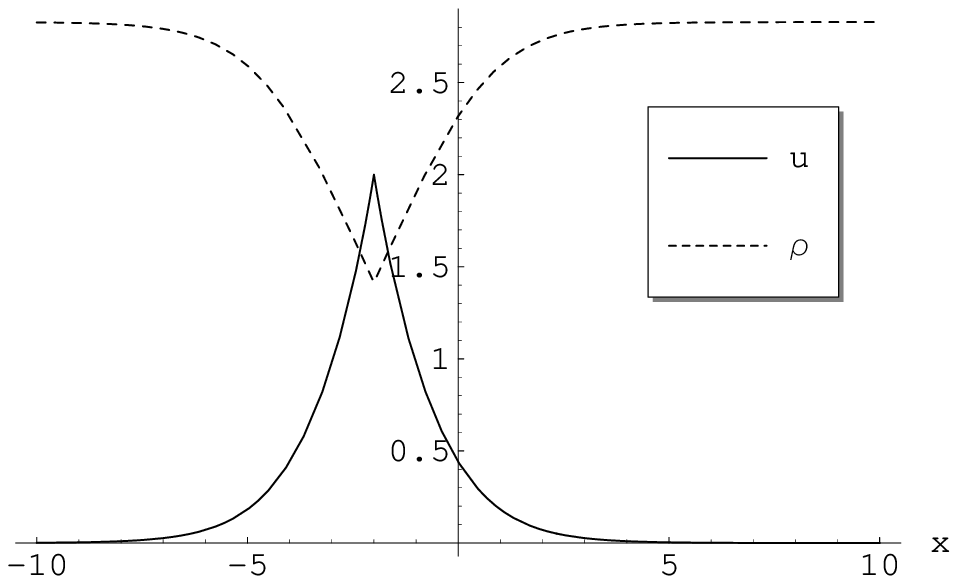} &
\includegraphics[width=0.45\textwidth]{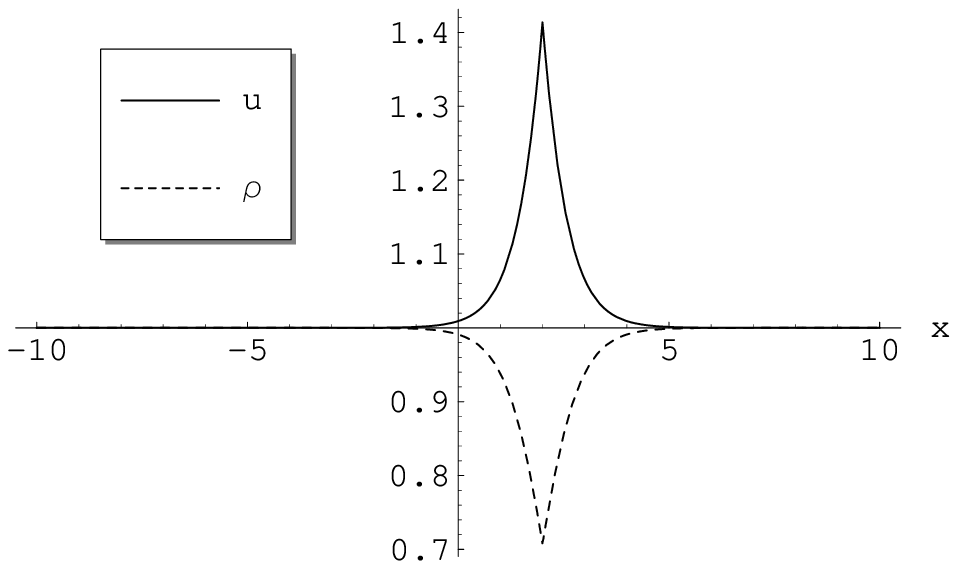} \\
\mbox{(a) (\ref{pk-1}) with } v=2,K=1,t=1. &
\mbox{(b) (\ref{pk-2}) with } p_1=1,x_0=2.
\end{array}
$$
\end{center}
\caption{Travelling peakon and stationary peakon}
\end{figure}

{\bf Example 3.} In this example we give some explicit expressions of kink and 2-kink interaction solutions of
the 2-CH system with $\kappa=0$. These solutions
are derived from the particular trivial solutions of the first negative flow of the AKNS hierarchy
given in the end of the last section and by using the B\"acklund transformations of Corollary \ref{bkl}. It's easy to
see that there are only constant
solutions when $n=1$. We consider here the cases when $n=2$ and $n=3$.

Let us first assume $n=2$. Denote  $\xi_i=p_i\,y+\frac{s}{p_i} +q_i$, then we have the following solution
for the first negative flow of the AKNS hierarchy
\begin{equation}
q=p_1 e^{\xi_1}+p_2 e^{\xi_2},\ r=0,\ b=2(e^{\xi_1}+e^{\xi_2}),\ c=0,\ a=1,
\end{equation}
where $p_1 \ne p_2$. By applying Theorem \ref{qrf} and the B\"acklund transformations of Corollary \ref{bkl} we arrive
at the following two primary solutions of the 2-CH system
\begin{equation}
f_0=\log\left(e^{\xi_1}+e^{\xi_2}\right),\
f_1=\log\left(\frac{(p_1-p_2)^2 e^{\xi_1+\xi_1}}{p_1^2 e^{\xi_1}+p_2^2 e^{\xi}_2}\right).
\end{equation}
Note that a further application of the B\"acklund transformations of Corollary \ref{bkl} leads to
$f_2=\log(0)$. So in this case we can only obtain two primary solutions. By using Theorem \ref{fx} we obtain
two solutions of the 2-CH system which have the form (\ref{zh8}) with the function $x(y,s)$ given respectively by
\begin{equation}\label{kink}
x_0=\log\left(e^{\tilde{\xi}_1}+e^{\tilde{\xi}_2}\right),\
x_1=\log\left(\frac{(p_1-p_2)^2 e^{\tilde{\xi}_1+\tilde{\xi}_1}}{p_1^2 e^{\tilde{\xi}_1}+p_2^2 e^{\tilde{\xi}_2}}\right).
\end{equation}
where $\tilde{\xi}_i=p_i\,s+\frac{y}{p_i}+q_i$. The solution obtained from $x_0$ (respectively, $x_1$) is an antikink
(respectively, kink), see Figure 2 where the profiles of $\rho$ are represented by the dashed curves.

\begin{figure}
\begin{center}\small
$$
\begin{array}{cc}
\includegraphics[width=0.45\textwidth]{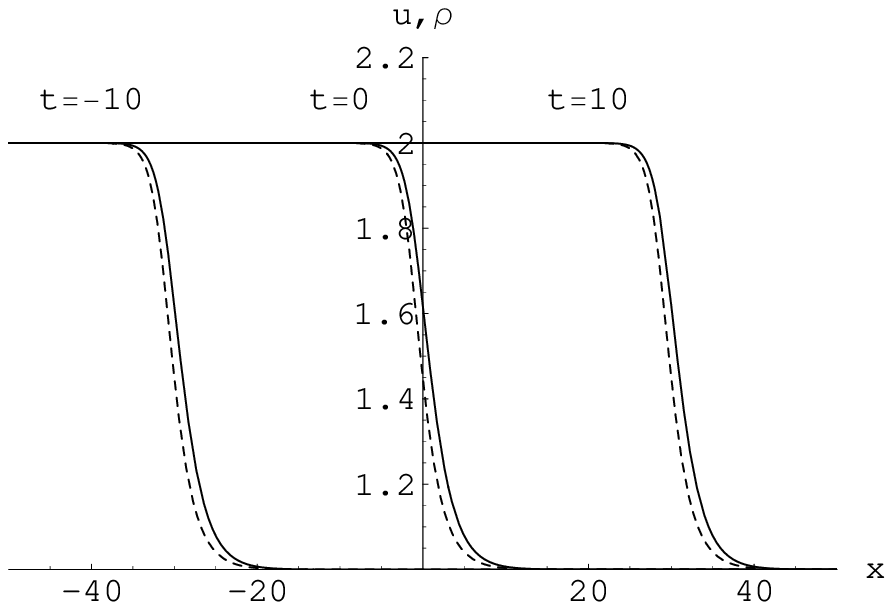} &
\includegraphics[width=0.45\textwidth]{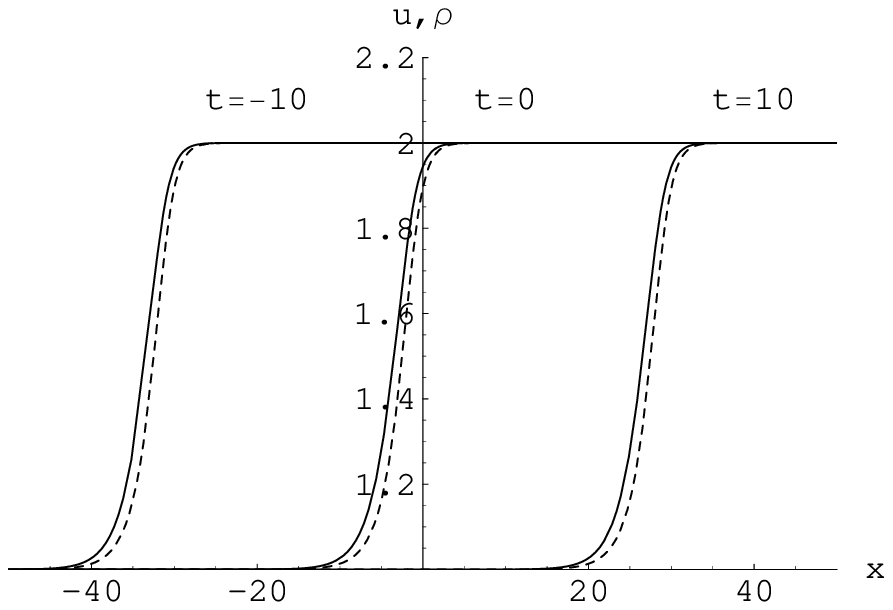} \\
\mbox{(a) } u,\rho \mbox{ obtained from } x_0. &
\mbox{(b) } u,\rho \mbox{ obtained from } x_1.
\end{array}
$$
\end{center}
\caption{Kink and antikink. ((\ref{kink}) with $p_1=1,p_2=2,q_1=0,q_2=0.$)}
\end{figure}

Now let us consider the case when $n=3$, in a similar way to the previous case we obtain three
solutions of the 2-CH system which have the form (\ref{zh8}) with the function $x(y,s)$ given respectively by
\begin{eqnarray}
x_0&=&\log\left(e^{\tilde{\xi}_1}+e^{\tilde{\xi}_2}+e^{\tilde{\xi}_3}\right), \label{kk-x0}\\
x_1&=&\log\left(\frac{p_1^2(p_2-p_3)^2e^{\tilde{\xi}_2+\tilde{\xi}_3}+p_2^2(p_3-p_1)^2e^{\tilde{\xi}_3+\tilde{\xi}_1}
+p_3^2(p_1-p_2)^2e^{\tilde{\xi}_1+\tilde{\xi}_2}}{p_2^2p_3^2e^{\tilde{\xi}_1}+p_3^2p_1^2e^{\tilde{\xi}_2}+
p_1^2p_2^2e^{\tilde{\xi}_3}}\right), \label{kk-x1} \\
x_2&=&\log\left(\frac{(p_1-p_2)^2(p_2-p_3)^2(p_3-p_1)^2e^{\tilde{\xi}_1+\tilde{\xi}_2+\tilde{\xi}_3}}{
p_1^4(p_2-p_3)^2e^{\tilde{\xi}_2+\tilde{\xi}_3}+p_2^4(p_3-p_1)^2e^{\tilde{\xi}_3+\tilde{\xi}_1}
+p_3^4(p_1-p_2)^2e^{\tilde{\xi}_1+\tilde{\xi}_2}}\right), \label{kk-x2}
\end{eqnarray}
where $p_1, p_2, p_3$ are pairwise distinct.
These solutions describe the antikink-antikink, antikink-kink and kink-kink interactions, see Figure 3,4,5.
These figures are drawn with parameters
\begin{equation}
p_1=1,\ p_2=2,\ p_3=3,\ q_1=0,\ q_2=0,\ q_3=0
\end{equation}
and only for $u$, since the figures of $\rho$ corresponding to the above three solutions of
the 2-CH system are
very similar to that of $u$, like the figures of the single kink solutions drawn in Figure 2.

In general, for any $n\in\mathbb N$ we expect to arrive at in this way $n$ solutions of the 2-CH system, each of which
is a $(n-1)$-kink solution if we choose the parameters $p_i, q_i$ in an appropriate way. They correspond
to the interactions of $k$ antikinks and $n-1-k$ kinks with $k=0,\dots, n-1$.
Except for the cases of $n=2, 3$, we can also check
this assertion for the case when $n=4$. We will leave the analysis of the general case to a subsequent
publication.

\begin{figure}
\begin{center}
\includegraphics[width=0.45\textwidth]{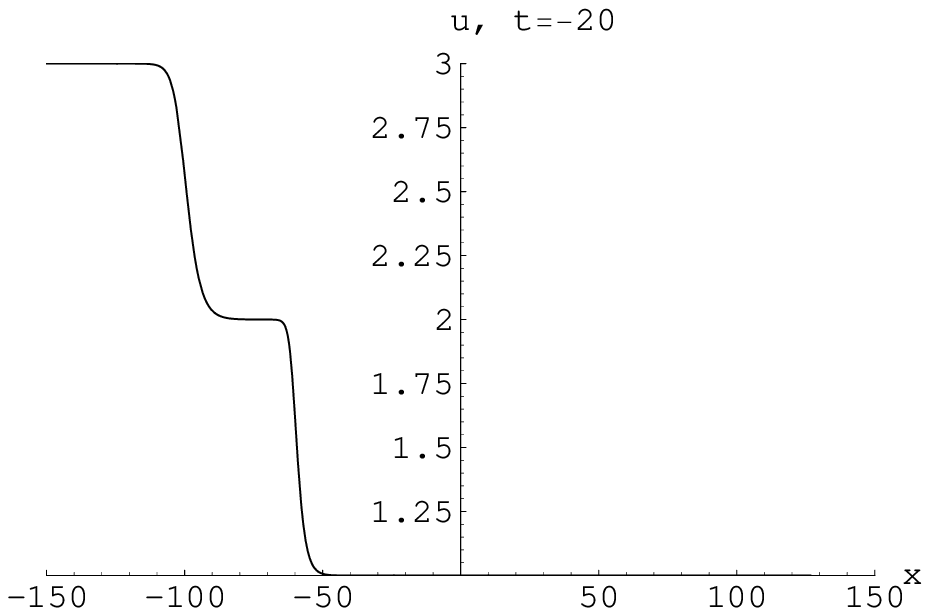}
\includegraphics[width=0.45\textwidth]{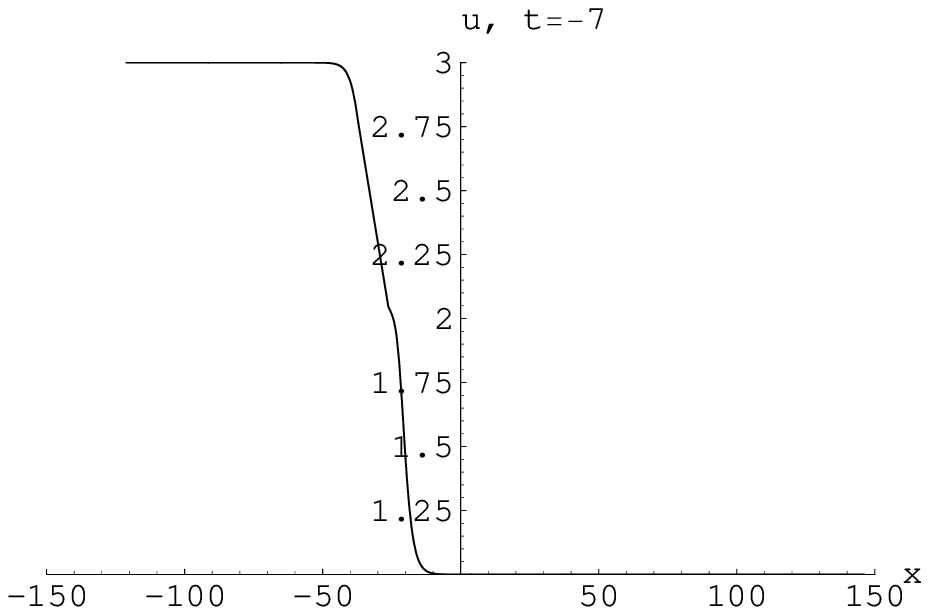}
\includegraphics[width=0.45\textwidth]{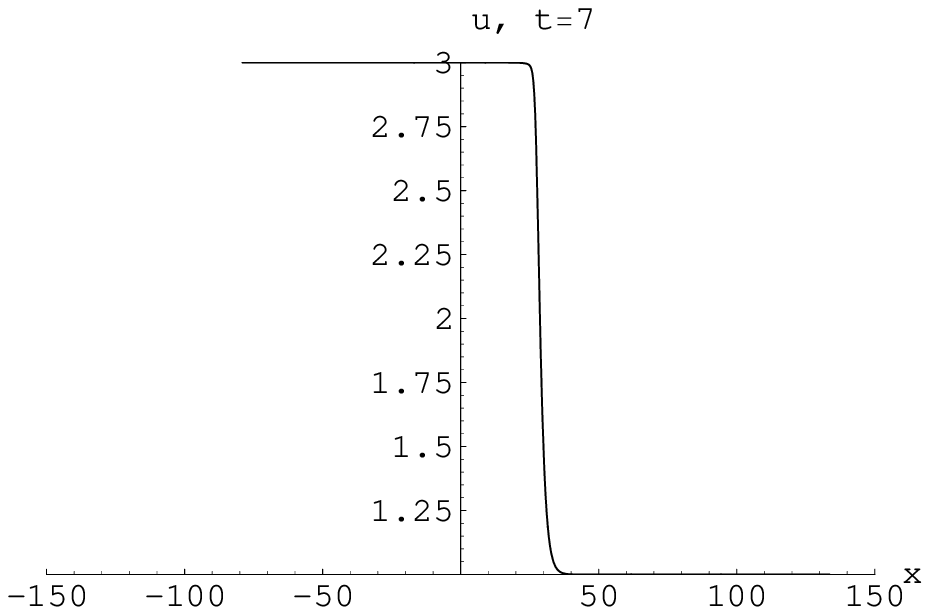}
\includegraphics[width=0.45\textwidth]{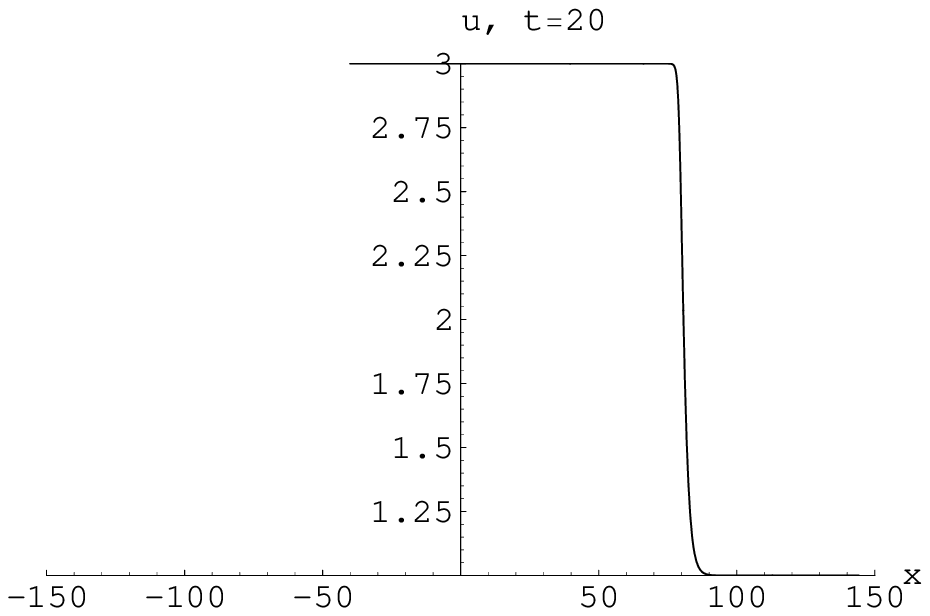}
\end{center}
\caption{The interaction of two antikinks (\ref{kk-x0}).}
\end{figure}

\begin{figure}
\begin{center}
\includegraphics[width=0.45\textwidth]{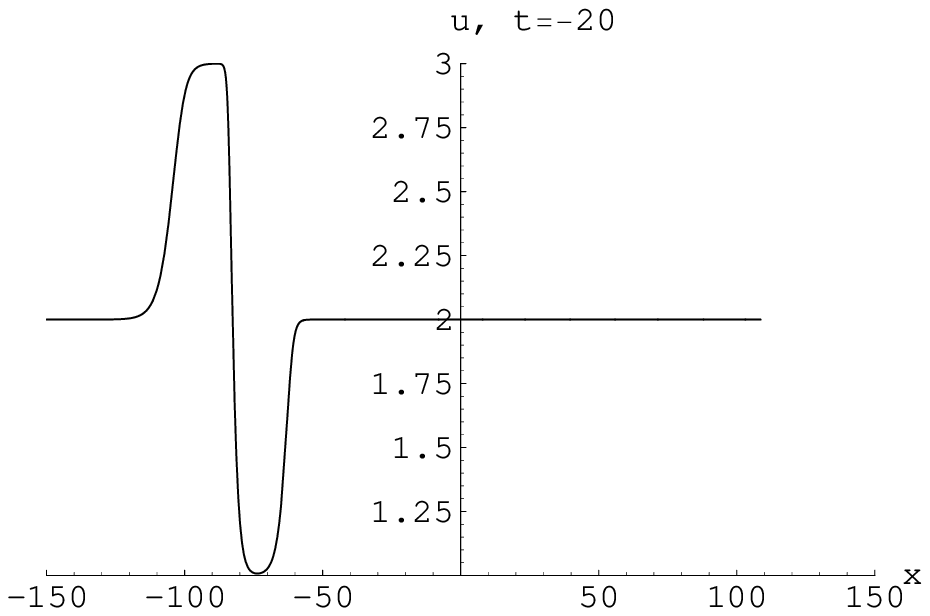}
\includegraphics[width=0.45\textwidth]{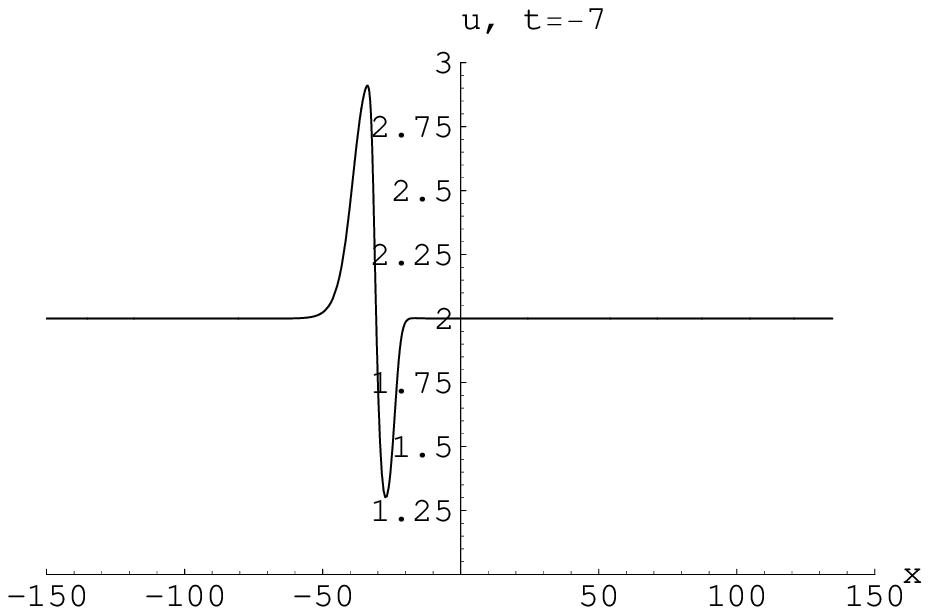}
\includegraphics[width=0.45\textwidth]{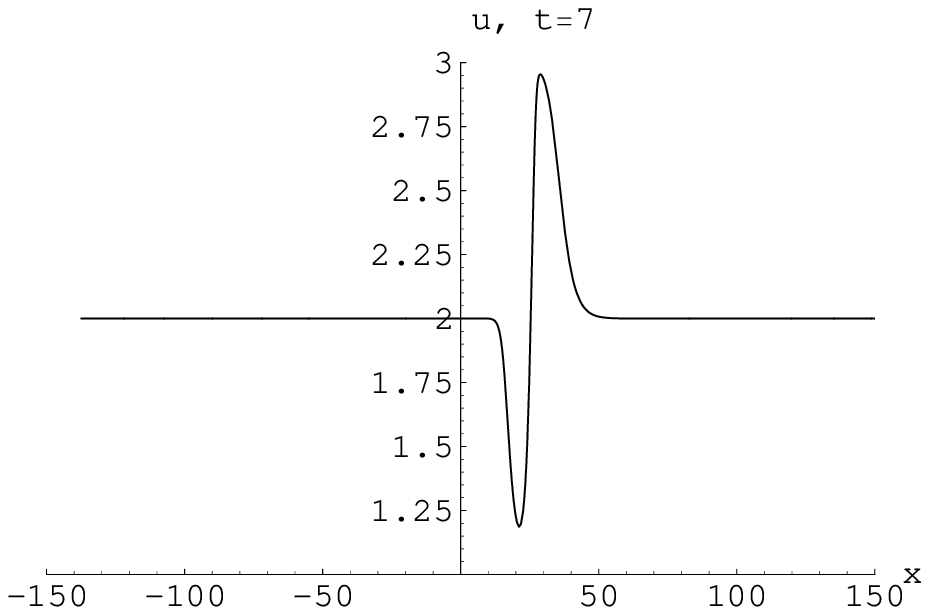}
\includegraphics[width=0.45\textwidth]{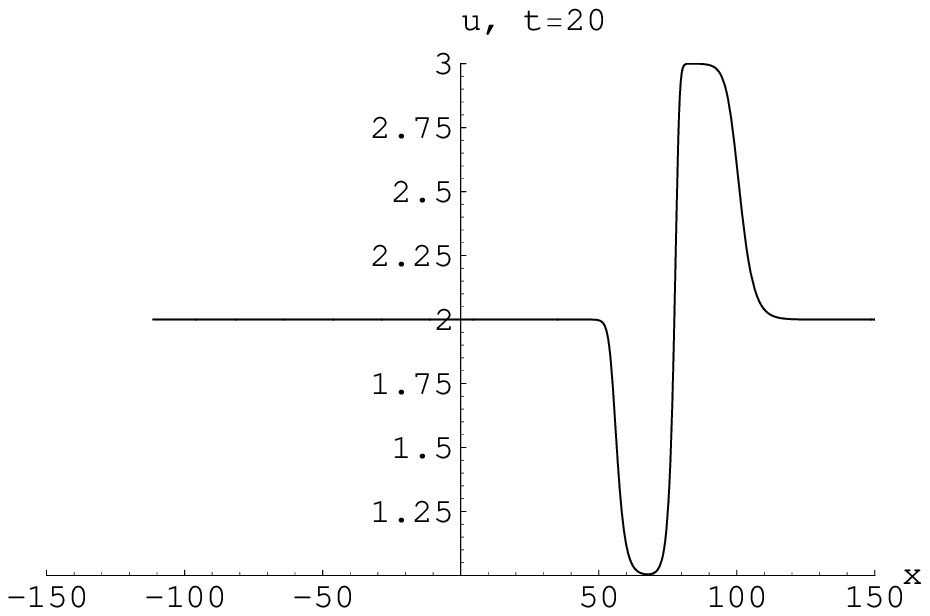}
\end{center}
\caption{The interaction of an antikink and a kink (\ref{kk-x1}).}
\end{figure}

\begin{figure}
\begin{center}
\includegraphics[width=0.45\textwidth]{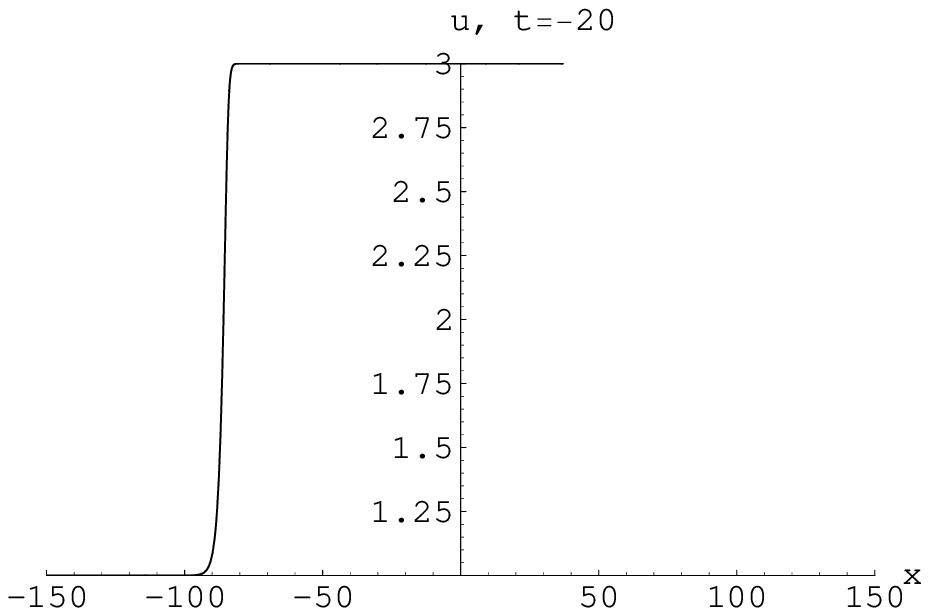}
\includegraphics[width=0.45\textwidth]{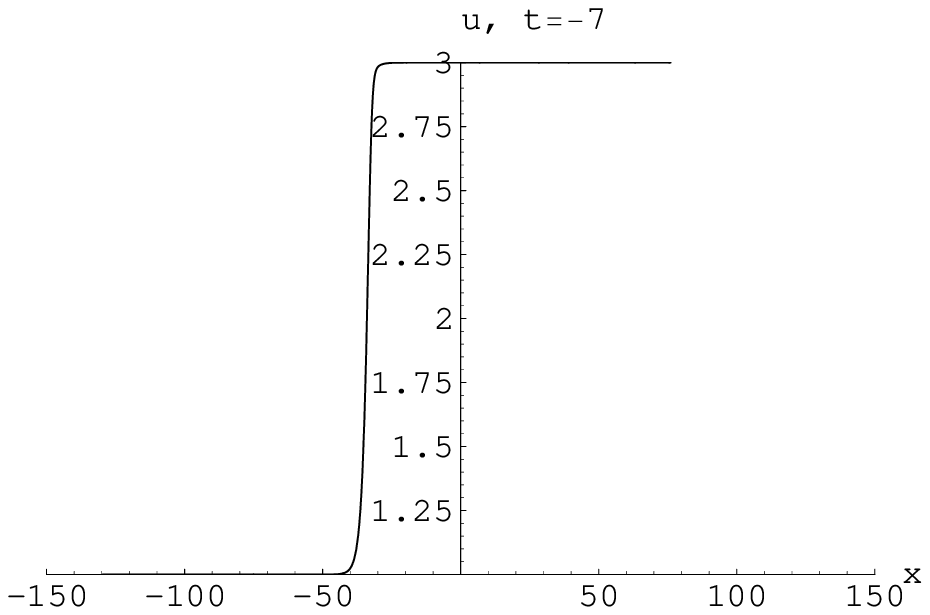}
\includegraphics[width=0.45\textwidth]{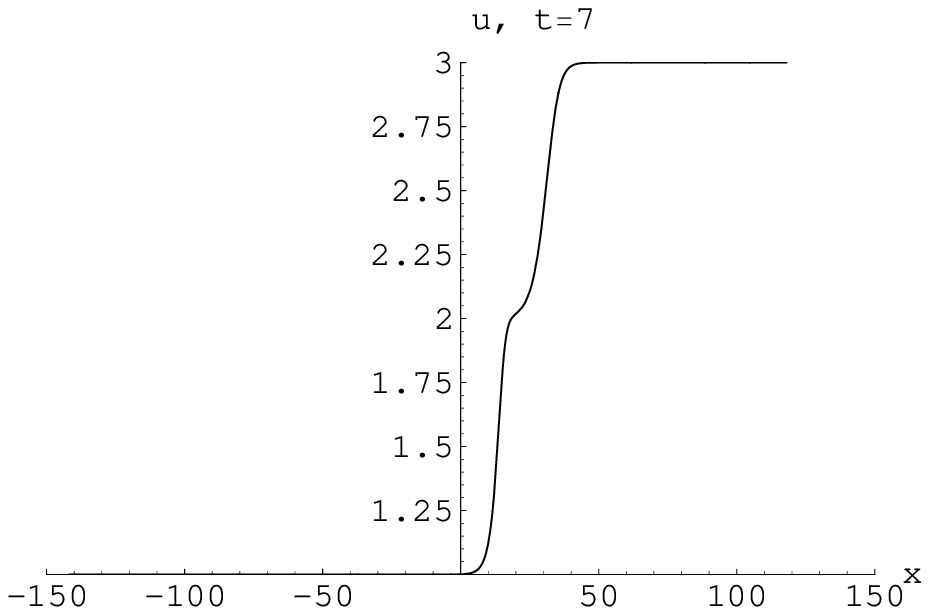}
\includegraphics[width=0.45\textwidth]{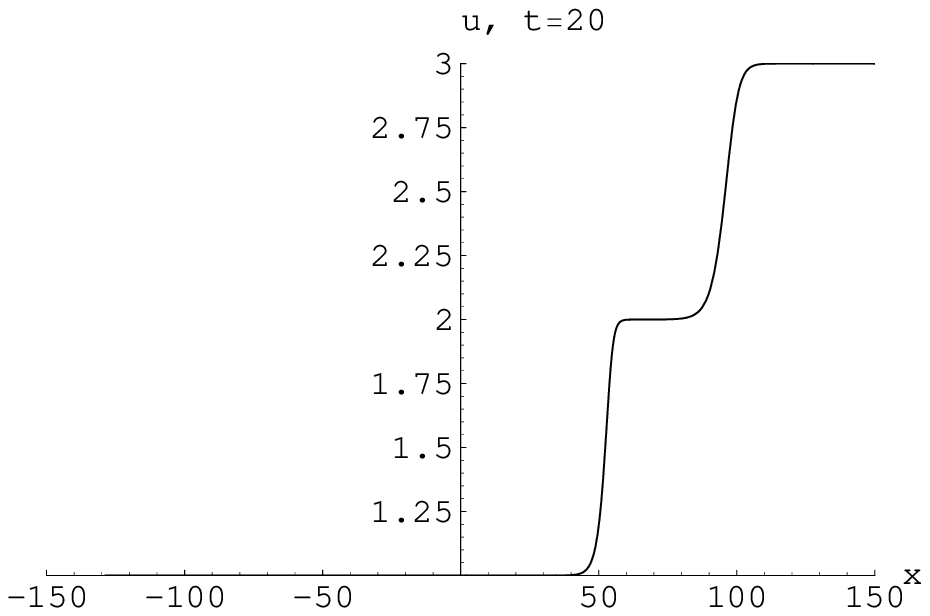}
\end{center}
\caption{The interaction of two kinks (\ref{kk-x2}).}
\end{figure}

\section{Conclusion}
We have constructed an explicit reciprocal transformation between the 2-CH system (\ref{ch-2}), (\ref{ch-3}) and the first negative
flow of the AKNS hierarchy (\ref{ak-1})--(\ref{ak-3}) with $\VE^2=1$. The primary solution $f(y,s)$ satisfying
the equation (\ref{ef})
plays a crucial role in this construction. This transformation comprises of two
step, the first step is the correspondence between solutions of the 2-CH system and the primary solutions
satisfying (\ref{ef}), it is given by the formulae (\ref{tr-2}), (\ref{tr-3}) and Theorems \ref{fuv}, \ref{fx}.
The second step is the correspondence between solutions of the first negative
flow of the AKNS hierarchy and the primary solutions of the 2-CH system, it is given by Theorems \ref{qrf}, \ref{fqr}.
These correspondences are presented in simple and explicit forms, they enable us to obtain solutions of the
2-CH system from that of the first negative flow of the AKNS hierarchy, which includes in particular the well
known sine-Grodon and the sinh-Gordon equations.

In terms of the primary solutions we also obtained
in Sec.\ref{sec-2}, \ref{sec-3} two kinds of B\"acklund transformations for the 2-CH system or
the first negative flow of the AKNS hierarchy, and showed in Sec.\ref{sec-4}
that the B\"acklund transformations given in this section
lead to interesting multi-kink solutions of the 2-CH system. It would be interesting to express
in terms of the primary solutions of the 2-CH system the
B\"acklund transformations of the AKNS hierarchy that are well known in the literatures, see for example \cite{gz,
matveev}.

We note that the travelling peakon solution of the 2-CH system that is given by
the first example of Sec.\ref{sec-4} assumes that the constant $K>0$, when $K=0$
it degenerates to a peakon solution of the Camassa-Holm equation (\ref{CH}).
The existence of multi-kink solutions of the 2-CH system is also
an intersting phenomenon for a system of hydrodynamic type with dispersive perturbation terms.
We will return to analyze in a subsequent paper in more details the various properties of particular solutions of the
2-CH system, including the problem of existence of multi-peakon solutions.
Although the
2-CH system that we considered here was derived from the problem of classification of  deformations of
bihamitionian structures of hydrodynamic type, we do expect that it would find for itself physically
important applications.

\vskip 0.2truecm \noindent{\bf Acknowledgments.} The authors are
grateful to Boris Dubrovin for helpful discussions and comments. They also
thank Yishen Li for drawing their attention to the interesting papers \cite{LZJ1, LZJ2}.
The researches of Y.Z. were partially
supported by the Chinese National Science Fund for Distinguished
Young Scholars grant No.10025101 and the Special Funds of Chinese
Major Basic Research Project ``Nonlinear Sciences''.

\eqa
{\rm email\ addresses:} && {\rm {\mbox{chen-02}}@mails.tsinghua.edu.cn,\ lsq99@mails.tsinghua.edu.cn} \nn\\
&& {\rm yzhang@math.tsinghua.edu.cn}\nn
\eeqa
\end{document}